**Subdivisions of the posteromedial cortex in disorders of consciousness**

**Running head:** Posteromedial cortex subdivision and consciousness

Yue Cui, PhD[1,2,3,*], Ming Song, PhD[1,2,3,*], Darren M. Lipnicki, PhD[4], Yi Yang, MD[5], Chuyang Ye, PhD[1,2,3], Lingzhong Fan, MD[1,2,3], Jing Sui, PhD[1,2,3,6,9], Tianzi Jiang, PhD[1,2,3,7,8,9,#], Jianghong He, MD[5,#]

[1] Brainnetome Center, Institute of Automation, Chinese Academy of Sciences, Beijing 100190, China

[2] National Laboratory of Pattern Recognition, Institute of Automation, Chinese Academy of Sciences, Beijing 100190, China

[3] University of Chinese Academy of Sciences, Beijing, China

[4] CHeBA (Centre for Healthy Brain Ageing), School of Psychiatry, University of New South Wales, Randwick, NSW 2031, Australia

[5] Department of Neurosurgery, PLA Army General Hospital, Beijing, China

[6] The Mind Research Network and Lovelace Biomedical and Environmental Research Institute, Albuquerque, NM 87106, USA

[7] Key Laboratory for NeuroInformation of Ministry of Education, School of Life Science and Technology, University of Electronic Science and Technology of China, Chengdu 610054, China

[8] Queensland Brain Institute, University of Queensland, Brisbane, QLD 4072, Australia

[9] CAS Center for Excellence in Brain Science and Intelligence Technology, Institute of Automation, Chinese Academy of Sciences, Beijing 100190, China

**Corresponding authors:**

[#] Jianghong He, MD, Department of Neurosurgery, PLA Army General Hospital, Beijing, China (Phone: +86-1-66721961; Fax: +86-1-64057752; E-mail: he_jianghong@sina.cn)

[#] Tianzi Jiang, PhD, Brainnetome Center, Institute of Automation, Chinese Academy of Sciences, Beijing 100190, China (Phone: +86-1-82544778; Fax: +86-1-82544777; E-mail: jiangtz@nlpr.ia.ac.cn)

[*] These authors contributed equally to the manuscript.

**Abstract**

Evidence suggests that disruptions of the posteromedial cortex (PMC) and posteromedial corticothalamic connectivity contribute to disorders of consciousness (DOCs). While most previous studies treated the PMC as a whole, this structure is functionally heterogeneous. The present study investigated whether particular subdivisions of the PMC are specifically associated with DOCs. Participants were DOC patients, 21 vegetative state/unresponsive wakefulness syndrome (VS/UWS), 12 minimally conscious state (MCS), and 29 healthy controls. Individual PMC and thalamus were divided into distinct subdivisions by their fiber tractograpy to each other and default mode regions, and white matter integrity and brain activity between/within subdivisions were assessed. The thalamus was represented mainly in the dorsal and posterior portions of the PMC, and the white matter tracts connecting these subdivisions to the thalamus had less integrity in VS/UWS patients than in MCS patients and healthy controls, as well as in patients who did not recover after 12 months than in patients who did. The structural substrates were validated by finding impaired functional fluctuations within this PMC subdivision. This study is the first to show that tracts from dorsal and posterior subdivisions of the PMC to the thalamus contribute to DOCs.

# 1. Introduction

Patients surviving severe brain damage may develop a long-term disorder of consciousness (DOC), such as vegetative state/unresponsive wakefulness syndrome (VS/UWS) or minimally conscious state (MCS). Evidence from brain imaging studies suggests that disconnections within thalamocortical areas of the default mode network (DMN) are implicated in DOCs (Boly et al., 2009; Fernández-Espejo et al., 2011; Fernandez-Espejo et al., 2012; He et al., 2015; Rosazza et al., 2016; Soddu et al., 2012; Vanhaudenhuyse et al., 2010). The posteromedial cortex (PMC, i.e., precuneus/posterior cingulate cortex) is the structural (Hagmann et al., 2008) and functional (Utevsky et al., 2014) core of the DMN, and diffusion magnetic resonance imaging (dMRI) of DOC patients has revealed white matter damage in connections between the PMC and thalamus (Fernandez-Espejo et al., 2012). A resting state functional magnetic resonance imaging (fMRI) study found reduced PMC-thalamic fluctuations in DOC patients compared with healthy controls (Boly et al., 2009), which accords well with our previous findings of disrupted functional connectivity between the PMC and thalamus in DOC patients (He et al., 2015). These findings suggest that the PMC and thalamus play important roles in determining levels of consciousness. Other studies have shown fMRI or PET can help in predicting recovery from a DOC. For example, functional connectivity strength discriminated between DOC patients who regained consciousness within 3 months and those who did not, with the most discriminative region being the PMC (Wu et al., 2015), and PET predicted long-term recovery (i.e., 12 months after baseline assessment) of patients with VS/UWS (Stender et al., 2014).

Previous studies of the PMC in DOC patients have investigated this large region as a whole (Boly et al., 2009; Fernandez-Espejo et al., 2012; He et al., 2015; Laureys et al., 1999a; Vanhaudenhuyse et al., 2010). However, recent work provides strong evidence of anatomical diversity and functional heterogeneity in the PMC (Margulies et al., 2009; Zhang et al., 2014). We previously used whole brain white matter tractography in healthy participants to reveal particularly strong connections between dorsal parts of the PMC and the thalamus (Zhang et al., 2014). Dorsal parts of the PMC are reported to have fundamental roles in self-awareness (Den Ouden et al., 2005; Kjaer et al., 2002) and in altered state of consciousness (Maquet et al., 1999), which suggests that, rather than the structure as a whole, particular parts of the PMC may be specifically involved in determining levels of consciousness. If so, this could have implications for using neuroimaging to facilitate more accurate diagnoses of DOCs and more reliably predict patient outcomes. The aim of the present study was to identify subdivisions of the PMC that may be specifically implicated in DOCs. We first used fiber tractography within regions of the DMN to identify subdivisions of the PMC and thalamus. White matter integrity between PMC and thalamus subdivisions, and brain activity within PMC subdivisions, was then compared across VS/UWS patients, MCS patients, and healthy controls, and between DOC patients who recovered after 12 months and those who did not.

## 2. Materials and methods

### 2.1 Participants

We recruited 79 patients with a DOC (VS/UWS, n = 62; MCS, n = 17) from the PLA Army General Hospital in Beijing between January 2014 and May 2016. All patients had received a severe brain injury caused by trauma, anoxia or stroke/cerebrovascular accident more than one month prior to recruitment. Patients were evaluated at least twice weekly within the two weeks before baseline, and their diagnosis of either VS/UWS or MCS was based on the Coma Recovery Scale-Revised (CRS-R) assessment with the highest diagnostic level during this time, as per Giacino et al. (2004). Exclusion criteria were having brain damage exceeding 30% of total brain volume, any contraindications to MRI scanning, being sedated or anesthetized during MRI acquisition, or having died during follow-up. The duration of follow-up was at least 12 months after baseline, with clinical examinations using the CRS-R and Glasgow Outcome Scale (GOS) (Jennett and Bond, 1975). Patients with a GOS level of 3-5 were considered to have recovered from their DOC (levels 1 and 2 represent dead and VS/UWS or MCS, respectively) The GOS has good inter-rater reliability of 92% (Wilson et al., 1998) and is commonly used to assess recovery from DOCs (e.g., Norton et al., 2012; Perlbarg et al., 2009; Qin et al., 2015; Van Der Eerden et al., 2014; Wu et al., 2015). The GOS has the advantage of being easy to administer and has a good inter-rater reliability of 92% (Wilson et al., 1998) that suggests the potential for bias across multiple raters is minimal. Most patients were evaluated at follow-up as outpatients, with those outside Beijing evaluated by local health care physicians. After excluding patients with incomplete follow-up assessments (n = 6), severe motion artifacts and poor quality MRI (n = 27), or failed registration or preprocessing because of severe structural deformations (n = 13), our study sample comprised 21 VS/UWS and 12 MCS patients. A group of 29 gender-matched healthy controls (HCs) were recruited. None of the HCs had a history of psychiatric or neurological illness, head injury, or drug or alcohol abuse. The study was approved by the Ethics Committee of the PLA Army General Hospital. All HCs and surrogates for the DOC patients provided written informed consent.

**2.2 MRI data acquisition**

All participants underwent dMRI, resting state fMRI and T1-weighted MRI scanning on a 3 T GE Discovery MR750 scanner (GE Medical Systems). The major acquisition parameters for dMRI data included: repetition time (TR)/echo time (TE) = 9000/81.9 ms, thickness/gap = 2.0/0 mm, flip angle = 90°, matrix size = 128 × 128 × 75, voxel size = 2 × 2 × 2 mm$^3$. For each participant, a total of 67 volumes were acquired, including 3 non-diffusion-weighted volumes (b = 0 s mm$^{-2}$) and 64 non-collinear gradient directions (b = 1000 s mm$^{-2}$). Resting state fMRI data were acquired axially using an echo-planar imaging sequence sensitive to blood oxygen level-dependent contrast. The acquisition parameters were TR/TE = 2000/30 ms, thickness/gap = 4.0/0.6 mm, flip angle = 90°, matrix size = 64 × 64 × 39, voxel size = 3.75 × 3.75 × 4

mm$^3$, 210 volumes were obtained for each participant. Sagittal T1-weighted structural MRI scans were obtained using the following optimized acquisition parameters: TR/TE = 8.16/3.18 ms, thickness/gap = 1.0/0 mm, flip angle = 7°, matrix size = 256 × 256 × 188; and voxel size = 1 × 1 × 1 mm$^3$.

**2.3 MRI data preprocessing**

The diffusion MR data were preprocessed using the FSL Diffusion Toolbox (FMRIB Software Library, http://fsl.fmrib.ox.ac.uk/fsl/fslwiki/FSL). First, corrections for eddy current distortions and head motion were performed by aligning all diffusion-weighted images to the non-diffusion-weighted image (the b0 volume). To ensure accurate brain masks and anatomical registration, T1-weighted images were processed with the Computational Anatomy Toolbox (CAT12, Structural Brain Mapping Group, http://dbm.neuro.uni-jena.de/cat/). The brain mask in diffusion space was then separated from the skull using the binarized skull-stripped T1-weighted images. Next, the skull-stripped T1-weighted image was aligned with b0 volume and diffusion images. In addition, each coregistered T1-weighted image was first linearly then nonlinearly warped to the Montreal Neurological Institute (MNI) template. The derived transformation parameters were then inverted and used to warp the seed and target masks from MNI space to the native dMRI space using a nearest neighbor interpolation. The registration procedures were performed using Aladin and F3D tools from the NiftyReg package (https://sourceforge.net/projects/niftyreg/). The diffusion-weighted data for each participant were visually inspected to ensure there were no apparent artifacts arising from acquisition or data processing procedures.

Resting state fMRI data were preprocessed in the native space of each participant using Data Processing & Analysis for Brain Imaging toolbox (DPABI) (Yan et al., 2016). Of the 210 volumes for each participant, the first 10 were discarded and the remainders underwent slice timing and motion correction, and were resampled to a stereoscopic 3 mm$^3$. In order to retain the temporal structures of patients with large head displacement, images with maximum displacement in the cardinal direction > 3 mm, or maximum spin > 3° from the previous frame were treated as outliers and included as nuisance regressors (Demertzi et al., 2015). In subsequent functional statistical analysis, we included head motion indexed by mean framewise displacement (Power et al., 2012) as a confounding factor in the statistical models. Linear regression was performed to remove the influence of head motion, linear trends, white matter and CSF signals. Finally, data were temporally filtered to keep frequencies of 0.01-0.08 Hz, to reduce low-frequency drift and high-frequency noise. Images were smoothed with a Gaussian filter of full width at half maximum value of 6 mm.

**2.4 Structural connectivity-based parcellation of the PMC and thalamus to the DMN**

We investigated 6 ROIs: the PMC, thalamus, superior frontal gyrus (SFG), angular gyrus, hippocampus, and medial prefrontal cortex (MPFC). Our connectivity-based parcellation scheme is similar to that previously reported by Cunningham et al (Cunningham et al., 2017). The mask for the PMC was manually drawn on a template of standard MNI152 space, with a detailed definition of the boundaries described elsewhere (Zhang et al., 2014). Other masks were obtained either from FSL's Harvard-Oxford Cortical and Subcortical Structural probabilistic atlases, including those for the thalamus, hippocampus, SFG and angular gyrus, or from the Stanford FIND Lab functional ROI database, for the MPFC (Shirer et al., 2012).

In our structural connectivity-based parcellation, either the PMC or thalamus was the seed, and the remaining 5 ROIs were targets. Probabilistic tractography in FSL was used to determine the number of streamlines (out of 5000) from each voxel of the seed region that had a 50% or greater chance of reaching each target ROI. Each voxel in the seed was assigned to the target ROI that received the highest number of streamlines from that voxel.

**2.5 Structural and functional analysis of consciousness levels at baseline**

Probabilistic tractography was performed in each participant's diffusion space to define pathways between distinct PMC subdivisions to corresponding thalamic subdivision and DMN regions. Voxels comprising PMC subdivisions were set as probabilistic tractography seed points (i.e., head region), with the corresponding thalamic subdivision and other DMN ROIs set as the waypoint, termination, and classification target (i.e., tail region, and vice versa). To minimize the effect of spurious connections, mean weighted fractional anisotropy (FA) values for the obtained path connecting head and tail were used to quantify path integrity. A voxel's weighted FA was calculated as the voxel's FA multiplied by the ratio of streamlines through the voxel to all streamlines in the path. Tractography is dependent on the seeding location, and the probability from the head to tail region is not necessarily equivalent to the probability from the tail to head region (Cao et al., 2013). For this reason, fiber tracking was performed in both directions between the ROIs, and the mean weighted FA is the average across these. For the resting state fMRI, we extracted the voxel-wise fractional amplitude of low-frequency fluctuations (fALFF) to generate a map for each subject as per Zou *et al.* (2008) using DPARSFA software. Individual fALFF maps were then registered into MNI standard space.

We analyzed main effects of group (i.e., VS/UWS, MCS, and HC) for mean weighted FA, with age and gender as statistical covariates. Functional validation was performed by comparing fALFF signals from the whole PMC between the VS/UWS and MCS groups. An AlphaSim approach (Ward, 2000) was used to correct for multiple comparisons. Potentially confounding factors controlled for in the functional analysis included age, gender, etiology, duration of illness, and mean framewise displacement movement factor.

**2.6 Structural and functional analysis in the prediction of recovery**

To investigate the prognostic capacity of our findings, we compared the non-recovered (GOS < 3) and recovered (GOS >= 3) groups on the structural integrity of tracts from the identified PMC subdivisions to the thalamus and other ROIs, and on PMC activity. For this we used general linear models that included age, gender, etiology and duration of illness as covariates. Statistical significance was set as a two-tailed $p < 0.05$.

## 3. Results

### 3.1 Demographic and clinical characteristics

Demographic and clinical characteristics of the participants are presented in Table 1 (with individual-level details for the patients in the Supplementary Table). There were no statistically significant differences in age, gender, or etiology between VS/UWS and MCS patients, though the VS/UWS patients had a shorter mean illness duration. At follow-up, 2 of 21 VS/UWS patients and 10 of 12 MCS patients were considered to have recovered (GOS >= 3).

### 3.2 Individual parcellation of the PMC and thalamus

The fine-grained architecture of the PMC and thalamus are shown by maximum probability maps (MPMs) and spaces for representative HC, MCS and VS/UWS individuals in Figure 1. We observed the thalamus to be represented mainly in the dorsal and posterior portions of the PMC (dpPMC), including the mesial extent of Brodmann areas 7 and 31, as well as area v23. The MPFC was represented mainly in the ventral and anterior parts of the PMC, and the PMC was represented mainly in regions of the thalamus that included the ventral posterior lateral nucleus (VPL), pulvinar (Pu) and parts of the centromedian nucleus (CM). The average streamlines from each subregion to corresponding target ROIs showed that structural connections decreased across the groups from HC to MCS to VS/UWS.

### 3.3 Structural and functional analysis of consciousness levels at baseline

As shown in Figure 2A, the tract between the dpPMC and VPL/Pu/CM mainly includes the splenium (posterior) and body of the corpus callosum, the posterior corona radiata, and retrolenticular part of the internal capsule. The mean weighted FA between the dpPMC and VPL/Pu/CM was significantly different among VS/UWS, MCS, and HC groups after controlling for age and gender ($F = 56.64$, $p < 0.001$, Figure 3B), with post hoc analyses showing that both the VS/UWS and MCS patient groups had less FA than the HC group ($p < 0.001$), and that VS/UWS patients had less FA than MCS patients ($p = 0.017$, Bonferroni corrected value). In addition, we found VS/UWS patients continued to have less FA than MCS patients ($p = 0.036$) when etiology and duration of illness were included as covariates in addition to age and gender. We also investigated whether the tracts between the ventral and anterior subdivisions of the PMC and the MPFC differed between groups. These tracts showed a significant group effect, with post-hoc analyses revealing differences between both patient groups and the HC group ($p < 0.05$), but not between the VS/UWS and MCS groups. Results of fMRI analyses revealed that VS/UWS patients had significantly lower fALFF values than MCS patients in the dorsal and ventral posterior PMC (AlphaSim corrected $p < 0.05$), with the peak location comprising 68.5% of all voxels being in the dpPMC (Figure 3). The remaining voxels were located in PMC subdivisions connecting to the MPFC (22.3%) and hippocampus (9.2%).

**3.4 Structural and functional analysis in the prediction of recovery**

Compared to the recovered group (GOS >= 3), the non-recovered group (GOS < 3) had less mean FA between the dpPMC and VPL/Pu/CM ($p = 0.031$). No significant effects were found for tracts between the ventral and anterior PMC and MPFC. The fALFF values were significantly decreased in the non-recovered group (AlphaSim corrected $p < 0.05$, Figure 4B); the peak voxel and 81.4% of the voxels showed reduced fALFF in the dpPMC. The remaining voxels were located in PMC subdivisions connecting to the MPFC (11.8%) and hippocampus (6.8%). Specifically, the fALFF values were significantly greater in recovered than in non-recovered VS/UWS patients (Figure 4C). To further validate our individual-level parcellation scheme, we used the MPM derived from our HCs, and a publicly available Brainnetome Atlas (Fan et al., 2016) with corresponding subregions of the dpPMC, and projected this into individual space. The results were unchanged from those found using our individualized parcellation scheme.

**4. Discussion**

To the best of our knowledge, this is the first study to show that tracts connecting specific PMC and thalamic subdivisions are implicated in DOCs. We found that dorsal and posterior portions of the PMC had pathways to the thalamus (VPL, Pu, and CM) with less white matter integrity, and had less resting state brain activity, in VS/UWS patients than in MCS patients, as well as in patients who did not recover after 12 months than in patients who did.

There being less PMC activity in VS/UWS patients than in MCS patients is consistent with previous work (Vanhaudenhuyse et al., 2010), and with evidence suggesting that the PMC plays a central role in mediating activity within the DMN (Fransson and Marrelec, 2008). The PMC has been previously reported as the region where activity is most predictive of whether coma and VS/UWS patients wake within 3 months (Wu et al., 2015), and seems to have a crucial role in the recovery of consciousness (Laureys et al., 1999b; Voss et al., 2006) that includes being the first cerebral region to increase activity in conscious waking from a vegetative state (Andreasen et al., 1995; Laureys et al., 1999a). A mesocircuit model of consciousness posits that cortico-thalamic-cortical loop systems involving the precuneus/posterior medial parietal complex are particularly vulnerable to injury (Giacino et al., 2014; Schiff, 2010) and that a progressive increase in neural activity within these networks leads to regaining consciousness (Laureys and Schiff, 2012). Our findings support the notion that disruptions within the corticothalamic system, particularly involving distinct subdivisions of the PMC and thalamus, underpin DOCs and the emerging from these. The PMC subdivisions we identified include the dorsal and posterior precuneus and ventral posterior cingulate cortex (PCC) posterior to the splenium of the corpus callosum. Previous research has shown important roles for these regions in various states of consciousness and conscious experience (Kjaer et al., 2002; Northoff and Bermpohl, 2004), and functional connectivity between the ventral PMC and MPFC is reported to predict recovery from coma (Silva et al., 2015). Our study extends the known roles of these regions in consciousness, by showing their involvement in DOCs.

The presence of dense connections between dorsal and posterior parts of the PMC and the thalamus in HCs has been previously reported (Behrens et al., 2003; Cunningham et al., 2017), and suggests the capacity of these tracts to convey rich information flow consistent with a role in the neural network subserving consciousness (Boly et al., 2009; Fernandez-Espejo et al., 2012). It is thus not surprising that we found disrupted connectivity between the PMC and thalamus in DOC patients and that this was greater for those with VS/UWS than with MCS, supporting previous reports of the structural integrity of these tracts being correlated with CRS-R scores (Vanhaudenhuyse et al., 2010; Wu et al., 2015), DOC categories (Fernandez-Espejo et al., 2012), and the recovery of consciousness from VS/UWS (Laureys et al., 2004). Others have also implicated tracts between the PMC and thalamus in the pathophysiologic basis of DOCs (Laureys and Schiff, 2012), and in the recovery from these (Koenig et al., 2014; Norton et al., 2012; Wu et al., 2015). Recovery from a vegetative state is also more likely with less damage to the corpus callosum (Kampfl et al., 1998), through which many of the PMC to thalamus tracts pass. These previous reports on recovery from DOCs fit with our finding of less PMC-thalamic disruption in DOC patients who recovered within 12 months than in those who did not. Some of our patients had traumatic head injuries, to which tracts between the PMC and thalamus are vulnerable (Leclercq et al., 2001) because of mechanical shearing forces (Gennarelli et al., 1982). However, while other of our DOC patients did not have a traumatic etiology, such cases have been previously reported to nevertheless have damage in white matter bundles including the corpus callosum (Molteni et al., 2017; Newcombe et al., 2010; Van Der Eerden et al., 2014) and thalamus (Newcombe et al., 2010).

We found that the tracts from specific subdivisions of the PMC targeted specific subdivisions of the thalamus, including the VPL/Pu/CM. Previous evidence also implicates the thalamus in DOCs. Thalamic lesions are common in VS/UWS patients (Adams et al., 1999), and more prevalent than in MCS patients (Jennett et al., 2001) in whom the relative preservation of corticothalamic connections might support consciousness (Fernández-Espejo et al., 2011) and residual cognitive functions (Giacino et al., 2014). It is also reported that deep brain stimulation of the bilateral centromedian-parafascicular complex can improve behavioral responsiveness and awareness in DOC patients (Chudy et al., 2017), and that a circuit-level mechanism involving the central thalamus and parietal regions helps regulate awareness (Giacino et al., 2014). Additionally, the thalamic subdivision we identified includes higher order nuclei rich in widely projecting matrix cells (Jones, 1998), the activity of which is thought to represent content-specific neural correlates of consciousness (Koch et al., 2016).

There are some limitations to the present study. The number of patients included in the analysis is relatively small, and the validity of our findings remains uncertain without replication in larger samples. A greater ratio of MCS patients to VS/UWS patients is also needed. It should also be noted that the generalizability of our study can apply only to patients able to be scanned without head movement artifacts, which can be especially present in DOC patients (Stender et al., 2014) because of limited or absent body control. We also did not study patients with severe structural deformation or damage in regions selected as ROIs. With only one follow-up we were not able to investigate relationships between neuroimaging markers and signs of transition from VS/UWS to MCS, and from MCS to normal consciousness. Having more follow-ups would also facilitate the development of an index or model for predicting the course and time to recovery.

The present study has shown the fine-grained architecture of the PMC and thalamus using diffusion tensor tractography, and revealed an involvement of dorsal and posterior sections of the posteromedial cortex, and the tracts between dpPMC and the VPL/Pu/CM region of the thalamus, in DOCs. This furthers our understanding of the neural basis of DOC, and offer potential avenues for treating these. Our findings may also casts light on the neurological mechanisms underpinning consciousness.

**Funding**

This work was partially supported by the Natural Science Foundation of China (Grant Nos. 31771076, 81771128, 91432302, 31620103905), the Science Frontier Program of the Chinese Academy of Sciences (Grant No. QYZDJ-SSW-SMC019), National Key R&D Program of China (Grant No. 2017YFA0105203), Beijing Municipal Science & Technology Commission (Grant Nos. Z141107002514111, Z161100000216152, Z161100000216139), and the Guangdong Pearl River Talents Plan (2016ZT06S220).

**Declaration of interest:**

All of the authors declare nothing to report.

**Figure legends**

**Figure 1.** The seeds and targets for the parcellation of the posteromedial cortex (PMC, A), and parcellation maps of the PMC (B) and the thalamus (C). First row shows the maximum probabilistic map for PMC and thalamic subdivisions and ROIs from a group of healthy controls (HCs) registered to Montreal Neurological Institute space. Second to fourth rows show parcellation maps and ROIs in individual space for a representative HC, minimally conscious state (MCS) patient, and a vegetative state / unresponsive wakefulness syndrome (VS/UWS) patient. Each voxel of the PMC (or thalamus) was assigned to 1 of 5 target DMN ROIs, which include the thalamus (or PMC), MPFC, hippocampus, superior frontal gyrus, and angular gyrus, based on the number of white matter connections to that region. Right column in B and C show structural connectivity fingerprints of all participants from HC, MCS, and VS/UWS groups showing the average number of streamlines from subregions connecting to corresponding targets. Red lines are the average number of streamlines for each group. DMN, default mode network; dpPMC, dorsal and posterior portions of the posteromedial cortex; MPFC, medial prefrontal cortex; SFG, superior frontal gyrus; VPL/Pu/CM, ventral posterior lateral nucleus, pulvinar and centromedian nucleus.

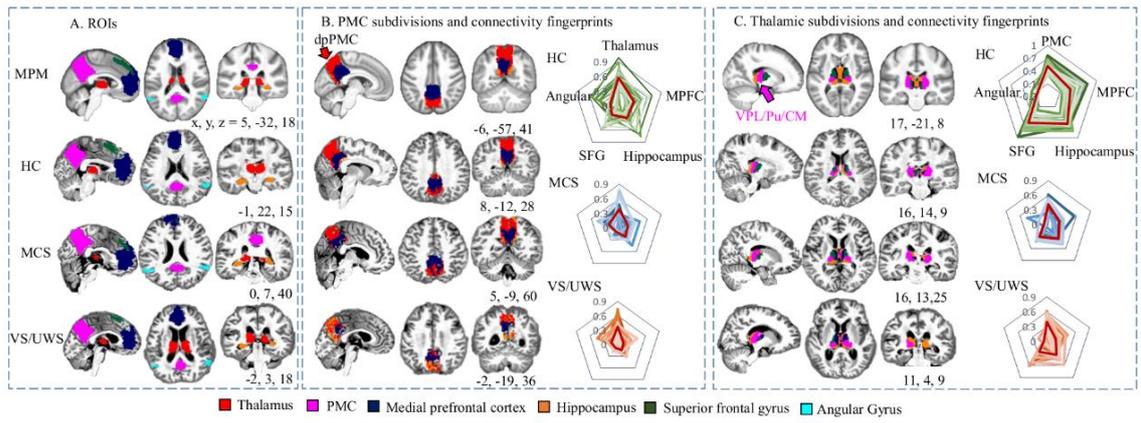

**Figure 2.** Structural connectivity analysis of consciousness levels at baseline. A, group probability maps of reconstructed tracts in healthy controls (HCs) between dpPMC (dorsal and posterior parts of the posteromedial cortex) and VPL/Pu/CM (ventral posterior lateral nucleus, pulvinar and center median nucleus). Maps are thresholded at presence in at least 25% of the participants. B, The mean weighted fractional anisotropy (FA) between dpPMC and VPL/Pu/CM was significantly different among VS/UWS, MCS and HC groups, controlling for the effects of age and gender, and with Bonferroni correction for multiple comparisons. ***, $p < 0.001$; *, $p < 0.05$.

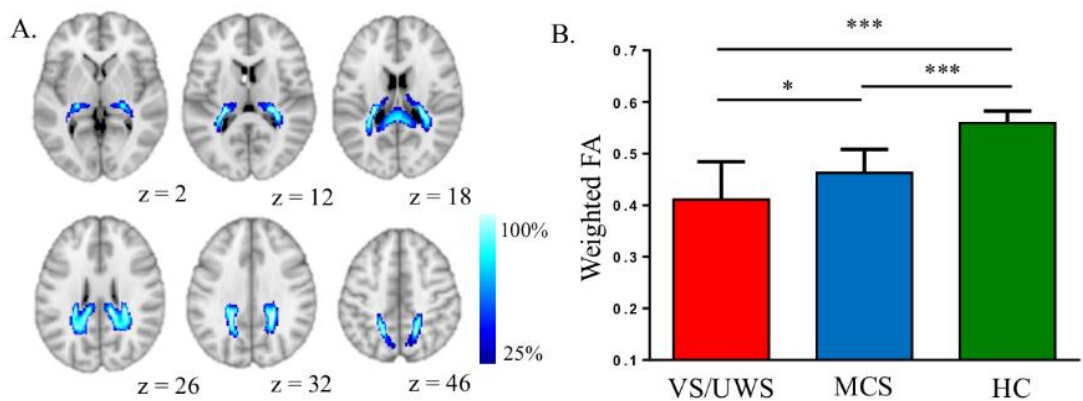

**Figure 3.** Functional activity analysis between vegetative state / unresponsive wakefulness syndrome (VS/UWS) and minimally conscious state (MCS) patients at baseline. A, Patients with VS/UWS showed significantly lower fractional amplitude of low-frequency fluctuations (fALFF) signal values than MCS patients (corrected *p* < 0.05), with the peak abnormal voxel and 68.5% of abnormal voxels in the subregion connecting to the thalamus (i.e., dorsal and posterior portions of the PMC, B and C).

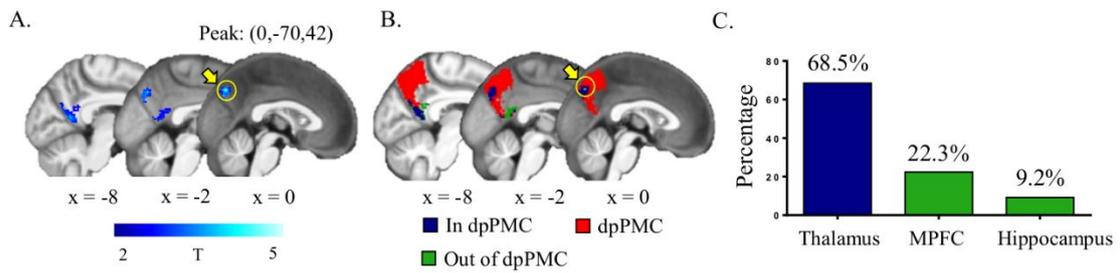

**Figure 4.** Structural and functional analysis in the prediction of recovery. A, The mean weighted fractional anisotropy (FA) was significantly lower in the non-recovered group (GOS < 3) than the recovered group (GOS >= 3). B, The fractional amplitude of low-frequency fluctuations (fALFF) signal values were significantly lower in the non-recovered group than the recovered group, with the peak abnormal voxel and 81.4% of abnormal voxels in the subregion connecting to the thalamus. C, The fALFF values were significantly greater in recovered than in non-recovered VS/UWS groups. Error bars represent SD. *, $p < 0.05$.

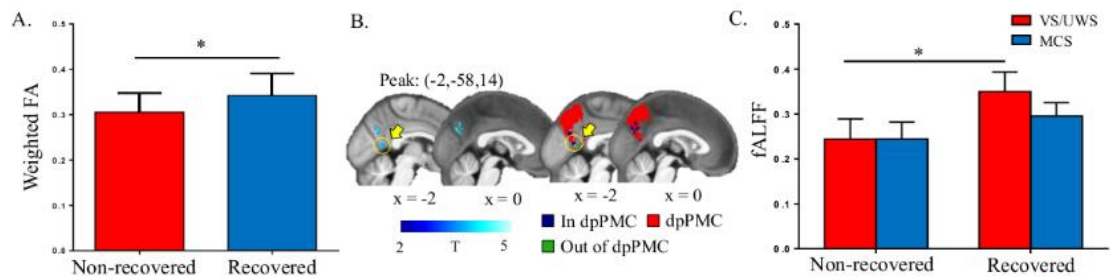

**Table 1**: Demographic and clinical characteristics of the participants

|  | VS/UWS<br>N = 21 | MCS<br>N = 12 | HC<br>N = 29 | Statistics<br>$F/\chi^2/t$ | $p$ |
|---|---|---|---|---|---|
| Age, years | 45.43 (14.33) | 37.25 (12.76) | 37.24 (11.27) | $F = 2.9$ | n.s. |
| Gender (male : female) | 8 : 13 | 8 : 4 | 15 : 14 | $\chi^2 = 1.27$ | n.s. |
| Etiology (TBI : CVA : ABI) | 7 : 5 : 9 | 6 : 5 : 1 |  | $\chi^2 = 1.73$ | n.s. |
| Duration of illness, months | 3.38 (1.72) | 14.42 (23.36) |  | $t = -2.18$ | 0.04 |
| Baseline CRS-R total | 6.05 (1.12) | 10.92 (3.85) |  | $t = -5.47$ | < 0.001 |
| Time interval between baseline and follow-up, months | 21.14 (8.37) | 18.42 (12.83) |  | $t = 0.74$ | n.s. |
| Follow-up CRS-R total | 8.76 (4.40) | 17.42 (4.21) |  | $t = -5.52$ | < 0.001 |
| Recovered |  |  |  |  |  |
|     Yes (GOS >= 3), N | 2 | 10 |  |  |  |
|     No (GOS < 3), N | 19 | 2 |  |  |  |

ABI, Anoxic brain injury; CRS-R, Coma Recovery Scale-Revised; CVA, stroke/cerebrovascular accident; GOS, Glasgow Outcome Scale; HC, healthy controls; MCS, minimally conscious state; MRI, magnetic resonance imaging; n.s., not significant; TBI, traumatic brain injury; VS/UWS, vegetative state/unresponsive wakefulness syndrome.

Mean and SD are reported unless otherwise specified.

**Supplementary Table.** Detailed demographic and clinical characteristics of patients

| Patient | Gender | Age, y | Group | Etiology | Duration, time prior to MRI, m | Follow-up, m | Structural lesions |
|---|---|---|---|---|---|---|---|
| 1 | M | 39 | MCS | ABI | 1 | 12 | Brainstem damage |
| 2 | M | 23 | MCS | TBI | 5 | 12 | Right basal ganglia (caudate) damage |
| 3 | M | 23 | MCS | CVA | 3 | 13 | Diffuse cortical & subcortical atrophy |
| 4 | M | 23 | MCS | TBI | 6 | 13 | Diffuse cortical & subcortical atrophy |
| 5 | M | 61 | MCS | ABI | 2 | 12 | Bilateral temporo-parietal lobe damage |
| 6 | M | 29 | MCS | TBI | 9 | 39 | Bilateral temporo-parietal damage |
| 7 | M | 42 | MCS | ABI | 3 | 12 | Left basal ganglia damage |
| 8 | M | 53 | MCS | ABI | 7 | 14 | Diffuse cortical & basal ganglia (caudate) damage |
| 9 | F | 46 | MCS | TBI | 77 | 51 | Diffuse axonal injury |
| 10 | F | 47 | MCS | ABI | 47 | 12 | Right basal ganglia damage |
| 11 | F | 28 | MCS | TBI | 1 | 19 | Left frontotemporal lobe damage |
| 12 | F | 33 | MCS | TBI | 12 | 12 | Bilateral frontal lobe damage and atrophy |
| 13 | F | 64 | VS/UWS | ABI | 1 | 17 | Left thalamus, basal ganglia lesions |
| 14 | F | 26 | VS/UWS | ABI | 4 | 12 | Left basal ganglia damage |
| 15 | F | 60 | VS/UWS | CVA | 4 | 13 | Diffuse hyperintense anoxic cortical lesions |
| 16 | M | 61 | VS/UWS | TBI | 9 | 16 | Right frontotemporal lobe damage |
| 17 | M | 51 | VS/UWS | CVA | 3 | 28 | Diffuse cortical & subcortical atrophy |
| 18 | F | 38 | VS/UWS | CVA | 2 | 19 | Diffuse cortical & subcortical atrophy |
| 19 | F | 30 | VS/UWS | CVA | 2 | 38 | Bilateral basal ganglia damage |
| 20 | F | 58 | VS/UWS | TBI | 2 | 14 | Diffuse cortical & subcortical atrophy |
| 21 | F | 25 | VS/UWS | CVA | 3 | 36 | Diffuse cortical & subcortical atrophy |
| 22 | M | 39 | VS/UWS | ABI | 3 | 12 | Right basal ganglia damage and atrophy |
| 23 | M | 41 | VS/UWS | CVA | 2 | 13 | Diffuse cortical & subcortical atrophy |
| 24 | F | 53 | VS/UWS | ABI | 3 | 28 | Bilateral brainstem and midbrain damage |
| 25 | F | 35 | VS/UWS | CVA | 3 | 19 | Diffuse cortical & subcortical |
| | | | | | | | atrophy |
| 26 | M | 46 | VS/UWS | TBI | 4 | 13 | Right temporo-parietal damage |
| 27 | M | 39 | VS/UWS | TBI | 5 | 16 | Right basal ganglia (caudate) damage |
| 28 | M | 22 | VS/UWS | TBI | 3 | 27 | Left fronto-temporo-parietal lobe damage |
| 29 | M | 51 | VS/UWS | TBI | 3 | 17 | Right frontotemporal lobe damage |
| 30 | F | 69 | VS/UWS | ABI | 4 | 33 | Diffuse cortical & subcortical atrophy |
| 31 | F | 68 | VS/UWS | TBI | 6 | 17 | Diffuse axonal injury |
| 32 | F | 35 | VS/UWS | CVA | 3 | 27 | Bilateral basal ganglia damage |
| 33 | F | 43 | VS/UWS | CVA | 2 | 29 | Diffuse cortical & subcortical atrophy |

**Supplementary Table.** Continued

| Patient | Baseline | | | | | | |
|---|---|---|---|---|---|---|---|
| | CRS-R | Auditory Function Scale | Visual Function Scale | Motor Function Scale | Oromotor/Verbal Function Scale | Communication Scale | Arousal Scale |
| 1 | 17 | 3 | 4 | 5 | 1 | 1 | 3 |
| 2 | 7 | 1 | 0 | 3 | 1 | 0 | 2 |
| 3 | 10 | 2 | 3 | 2 | 1 | 0 | 2 |
| 4 | 9 | 1 | 3 | 2 | 1 | 0 | 2 |
| 5 | 11 | 1 | 3 | 4 | 1 | 0 | 2 |
| 6 | 18 | 3 | 5 | 5 | 1 | 1 | 3 |
| 7 | 7 | 1 | 0 | 3 | 1 | 0 | 2 |
| 8 | 11 | 3 | 3 | 2 | 1 | 0 | 2 |
| 9 | 11 | 2 | 2 | 2 | 2 | 1 | 2 |
| 10 | 8 | 1 | 1 | 3 | 1 | 0 | 2 |
| 11 | 15 | 3 | 3 | 5 | 1 | 0 | 3 |
| 12 | 7 | 1 | 0 | 2 | 2 | 0 | 2 |
| 13 | 7 | 1 | 1 | 2 | 1 | 0 | 2 |
| 14 | 6 | 1 | 0 | 2 | 1 | 0 | 2 |
| 15 | 6 | 1 | 0 | 2 | 1 | 0 | 2 |
| 16 | 6 | 1 | 0 | 2 | 1 | 0 | 2 |
| 17 | 7 | 1 | 1 | 2 | 1 | 0 | 2 |
| 18 | 6 | 1 | 1 | 1 | 1 | 0 | 2 |
| 19 | 4 | 0 | 0 | 2 | 0 | 0 | 2 |
| 20 | 3 | 0 | 0 | 2 | 1 | 0 | 0 |
| 21 | 5 | 1 | 0 | 2 | 0 | 0 | 2 |
| 22 | 7 | 1 | 1 | 2 | 1 | 0 | 2 |
| 23 | 5 | 0 | 0 | 2 | 1 | 0 | 2 |
| 24 | 5 | 1 | 1 | 2 | 1 | 0 | 0 |
| 25 | 6 | 1 | 0 | 2 | 1 | 0 | 2 |
| 26 | 6 | 1 | 0 | 1 | 2 | 0 | 2 |
| 27 | 7 | 1 | 1 | 2 | 1 | 0 | 2 |
| 28 | 7 | 1 | 1 | 2 | 1 | 0 | 2 |
| 29 | 7 | 1 | 1 | 2 | 1 | 0 | 2 |
| 30 | 6 | 1 | 0 | 2 | 1 | 0 | 2 |
| 31 | 7 | 1 | 1 | 2 | 1 | 0 | 2 |
| 32 | 7 | 2 | 1 | 1 | 1 | 0 | 2 |
| 33 | 7 | 1 | 1 | 2 | 1 | 0 | 2 |

**Supplementary Table.** Continued

| Patient | CRS-R | Auditory Function Scale | Visual Function Scale | Motor Function Scale | Oromotor/ Verbal Function Scale | Communication Scale | Arousal Scale | GOS | Recovered |
|---|---|---|---|---|---|---|---|---|---|
| | | | | | **Follow-up** | | | | |
| 1 | 19 | 4 | 4 | 5 | 1 | 2 | 3 | 3 | Yes |
| 2 | 11 | 2 | 2 | 3 | 2 | 0 | 2 | 2 | No |
| 3 | 21 | 4 | 5 | 5 | 2 | 2 | 3 | 4 | Yes |
| 4 | 19 | 4 | 4 | 4 | 2 | 2 | 3 | 3 | Yes |
| 5 | 16 | 2 | 2 | 3 | 1 | 1 | 2 | 3 | Yes |
| 6 | 22 | 4 | 5 | 6 | 2 | 2 | 3 | 4 | Yes |
| 7 | 19 | 4 | 1 | 6 | 3 | 2 | 3 | 3 | Yes |
| 8 | 12 | 3 | 3 | 2 | 1 | 2 | 3 | 3 | Yes |
| 9 | 13 | 3 | 3 | 2 | 2 | 1 | 2 | 2 | No |
| 10 | 13 | 2 | 2 | 2 | 2 | 1 | 2 | 3 | Yes |
| 11 | 22 | 4 | 5 | 6 | 2 | 2 | 3 | 4 | Yes |
| 12 | 22 | 4 | 5 | 5 | 3 | 2 | 3 | 4 | Yes |
| 13 | 11 | 2 | 3 | 3 | 1 | 0 | 2 | 2 | No |
| 14 | 6 | 1 | 0 | 2 | 1 | 0 | 2 | 2 | No |
| 15 | 6 | 1 | 0 | 2 | 1 | 0 | 2 | 2 | No |
| 16 | 8 | 1 | 2 | 2 | 1 | 0 | 2 | 2 | No |
| 17 | 7 | 1 | 1 | 2 | 1 | 0 | 2 | 2 | No |
| 18 | 23 | 4 | 5 | 6 | 3 | 2 | 3 | 5 | Yes |
| 19 | 7 | 0 | 2 | 2 | 1 | 0 | 2 | 2 | No |
| 20 | 3 | 0 | 0 | 2 | 1 | 0 | 1 | 2 | No |
| 21 | 6 | 1 | 1 | 2 | 0 | 0 | 2 | 2 | No |
| 22 | 7 | 1 | 1 | 2 | 1 | 0 | 2 | 2 | No |
| 23 | 5 | 0 | 0 | 2 | 1 | 0 | 2 | 2 | No |
| 24 | 7 | 1 | 1 | 2 | 1 | 0 | 2 | 2 | No |
| 25 | 8 | 1 | 1 | 2 | 2 | 0 | 2 | 2 | No |
| 26 | 6 | 1 | 0 | 1 | 2 | 0 | 2 | 2 | No |
| 27 | 11 | 3 | 2 | 2 | 1 | 1 | 2 | 2 | No |
| 28 | 15 | 3 | 3 | 4 | 1 | 2 | 2 | 2 | No |
| 29 | 16 | 3 | 3 | 5 | 1 | 1 | 3 | 3 | Yes |
| 30 | 7 | 1 | 0 | 2 | 2 | 0 | 2 | 2 | No |
| 31 | 9 | 1 | 3 | 2 | 1 | 0 | 2 | 2 | No |
| 32 | 9 | 2 | 3 | 1 | 1 | 0 | 2 | 2 | No |
| 33 | 8 | 2 | 0 | 2 | 1 | 1 | 2 | 2 | No |

ABI, Anoxic brain injury; CRS-R, Coma Recovery Scale-Revised; CVA, stroke/cerebrovascular accident; GOS, Glasgow Outcome Scale; MCS, minimally conscious state; TBI, traumatic brain injury; VS/UWS, vegetative state/unresponsive wakefulness syndrome.